\documentclass[lettersize,journal]{IEEEtran}
\usepackage{amsmath,amsfonts}
\usepackage{algorithmic}
\usepackage{algorithm}
\usepackage{array}
\usepackage[caption=false,font=normalsize,labelfont=sf,textfont=sf]{subfig}
\usepackage{textcomp}
\usepackage{stfloats}
\usepackage{url}
\usepackage{verbatim}
\usepackage{graphicx}
\usepackage{cite}
\usepackage{booktabs}
\usepackage{multirow}
\usepackage[table]{xcolor}
\definecolor{starecgray}{gray}{0.92}
\begin{document}
\bstctlcite{IEEEexample:BSTcontrol}

\title{Where Reasoning Matters: \\
Rethinking Latent Reasoning in Semantic ID-based Generative Recommendation}

\author{Shangxin Yang, $\text{Min Gao}^{*}$, Zongwei Wang, Junliang Yu
        % <-this % stops a space

\thanks{*Corresponding Author}

\thanks{- S. Yang, M. Gao, and Z. Wang are with Key Laboratory of Dependable Service Computing in Cyber Physical Society
(Chongqing University), Ministry of Education, China; the School of Big Data and Software Engineering, Chongqing University, Chongqing, China. Emails: \{zongwei, gaomin, guolinxin\}@cqu.edu.cn and yangshangxin@stu.cqu.edu.cn.}

\thanks{J. Yu is with the School of Information and Communication Technology, Griffith University, Australia. E-mail: junl.yu@outlook.com.}

}

\markboth{}%
{Shell \MakeLowercase{\textit{et al.}}: A Sample Article Using IEEEtran.cls for IEEE Journals}


\maketitle

\begin{abstract}
Semantic ID-based generative recommendation predicts an item by generating a short sequence of semantic ID tokens, where each token is produced autoregressively. Latent reasoning has recently been introduced to improve this process through additional hidden-state computation before each token decision. This raises a practical question: when one item is represented by a sequence of semantic ID tokens, should each token receive the same fixed number of latent refinement steps, or should these steps be allocated more effectively across positions? We study this question through position-wise information-gain (IG), which measures how much each semantic ID position reduces the uncertainty of the target item. We observe that earlier semantic ID positions usually provide higher information-gain, while later positions contribute less additional information. We further analyze that applying more refinement to high-IG positions tends to bring larger expected benefits. Based on this observation, we propose IBA, an Information-Gain Budget Allocation framework for semantic ID-based generative recommendation. IBA treats latent refinement steps as a limited computational resource and learns how to allocate them across semantic ID positions, assigning more refinement to informative positions and less to positions with smaller contribution. Experiments on multiple public datasets show that IBA consistently improves strong generative recommendation baselines and achieves a better accuracy--computation trade-off than fixed or poorly matched step allocations.

\end{abstract}


\section{Introduction}
\label{sec:introduction}
\IEEEPARstart{R}{ecommender} systems have become a critical infrastructure for personalized services on large-scale e-commerce platforms \cite{zhou2018deep,zhang2019deep}. Conventional recommendation pipelines often rely on multi-stage retrieval and ranking with independent item IDs. Although effective, this design can accumulate errors across stages \cite{wang2025learning,wei2024enhancing} and provides limited semantic sharing for sparse or newly introduced items \cite{rajput2023recommender,zhang2019deep}. Generative recommendation addresses these limitations by directly generating item identifiers conditioned on user behavior. In particular, semantic ID-based methods represent each item as a sequence of discrete tokens produced by vector quantization methods such as RQ-VAE~\cite{lee2022autoregressive}, allowing related items to share code patterns and improving generalization beyond independent item identifiers \cite{rajput2023recommender,zhou2025onerec}.

Recent progress in reasoning-capable language models \cite{fu2025think,guo2025deepseek,qiu2026bayesian,shi2025swireasoning} has inspired latent reasoning for generative recommendation \cite{lin2026bringing,zhang2025slow,guo2026s$^2$gr}. Instead of generating explicit reasoning text, latent reasoning performs additional computation in hidden space before committing each output token, providing a promising way to improve recommendation quality at inference time. For semantic ID generation, however, one item is predicted through multiple token decisions rather than a single output. This raises a practical question: \textbf{whether every semantic ID position needs the same number of refinement steps?}

To study this question, we first examine whether all semantic ID positions contribute equally to identifying the target item. In semantic ID-based generative recommendation, two intrinsic structures suggest that the answer may be no. First, RQ-VAE quantization produces hierarchical semantic IDs: earlier positions tend to encode dominant item information, while later positions mainly refine residual details. Second, the generation trie restricts each next-token prediction to prefix-valid candidates, making the candidate space much smaller at later positions.

\begin{figure}[!t]
\centering
\includegraphics[width=\columnwidth]{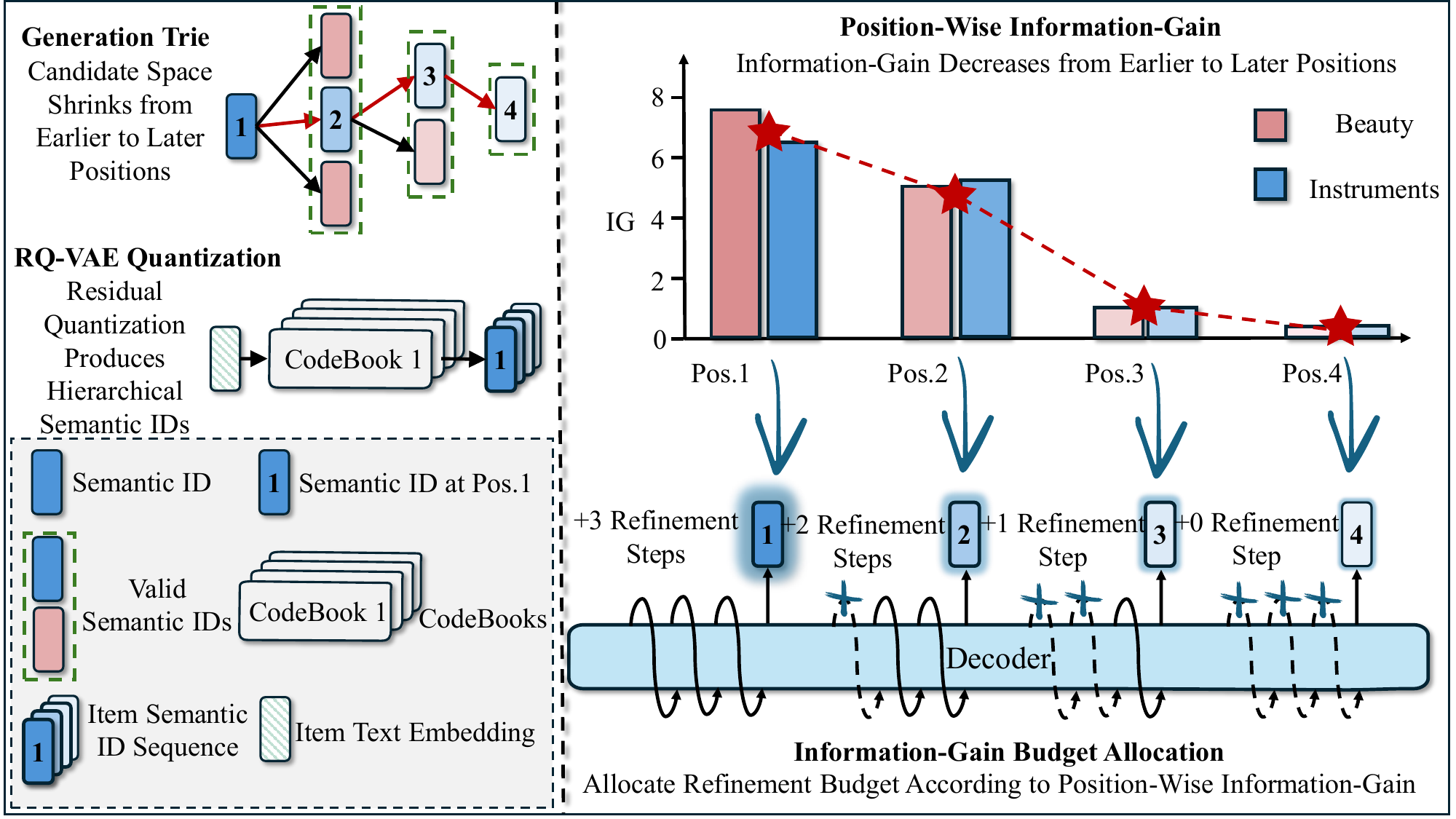}
\caption{Illustration of position-wise information-gain in semantic ID generation. Empirical IG on Instruments and Beauty shows a decreasing trend across positions, motivating IBA to allocate more refinement steps to high-IG early positions and fewer steps to low-IG later positions.}

\label{fig:rqvae_intro}
\end{figure}

These two structures jointly lead to a position-wise information-gain (IG) pattern, as visualized on the Instruments and Beauty datasets in Fig.~\ref{fig:rqvae_intro}. IG measures how much the uncertainty of the target item is reduced after observing the semantic ID token at a given position. Earlier positions usually reduce more uncertainty than later positions. However, this contribution matters only when the token is predicted correctly, which means a high-IG token can greatly reduce more ambiguity about the target item if it is correct, but a wrong prediction at this position also misdirects the whole semantic ID sequence.

Therefore, to evaluate whether a position needs one additional refinement step, we consider the benefit of that step from two factors: how much it improves the current token prediction, and how much uncertainty about the target item can be reduced once the token is correct. Supported by the empirical results in Section~\ref{subsec:ig_analysis}, our analysis shows that early positions tend to satisfy both conditions. They have larger IG, and additional refinement still improves their token-level accuracy, whereas later positions are already close to saturation under the trie constraint and receive little benefit from extra refinement.

This suggests that assigning the same number of latent refinement steps to every semantic ID position can waste computation on low-value decisions and underuse computation where refinement is more valuable. A more suitable strategy is to assign different numbers of latent refinement steps according to the expected refinement value of each position, using position-wise IG as a structural prior. However, this also raises two challenges. First, executing different numbers of refinement steps at different positions may make the reasoning trajectory unstable. Second, the position-wise IG pattern only describes the overall structure across items, while the benefit of an extra refinement step may still vary for each user sequence.

Motivated by this observation, we propose IBA, an Information-Gain Budget Allocation framework for semantic ID-based generative recommendation. Here, the budget refers to the number of latent refinement steps available during semantic ID generation. To address the instability caused by assigning different refinement steps according to the IG pattern, IBA introduces a Dual-Axis Refinement module. It performs semantic alignment along the vertical axis, allowing multi-step reasoning along the horizontal axis to proceed stably. To decide the allocation for each user sequence, IBA uses a lightweight refinement gain predictor to estimate the expected gain of candidate refinement steps and selects the allocation that maximizes the total gain under the overall IG pattern. To train this predictor, IBA adopts a two-stage training strategy that provides empirical supervision             signals for gain prediction. It further introduces a lookahead objective, encouraging high-IG early positions that receive more refinement steps to help predict later semantic ID positions, thereby increasing the utility of the limited budget.

We evaluate IBA on three public recommendation datasets using strong sequential, generative, and reasoning-enhanced baselines. The results show that IBA consistently improves semantic ID-based generative recommenders across datasets and backbones. Further analyses verify the position-wise IG pattern, compare learned and fixed step schedules, and examine the contribution of each model component.

The main contributions of this work are summarized as follows:
\begin{itemize}
\item We identify and quantify a position-wise IG pattern in semantic ID-based generative recommendation, showing that different code positions contribute unequally to item identification.

\item We propose IBA, a latent-reasoning framework that adaptively allocates latent refinement steps across semantic ID positions.

\item We design a two-stage training strategy that supports learned allocation of refinement steps across semantic ID positions.

 \item We conduct experiments on multiple public datasets, showing that IBA improves strong generative recommendation baselines and provides a better accuracy--computation trade-off than fixed-step or misaligned step allocations.
\end{itemize}

\section{Preliminaries}
\label{sec:preliminaries}
\label{subsec:preliminaries}

\subsection{Semantic ID-based Generative Recommendation}
Let $\mathcal{U}=\{u_1,u_2,\dots,u_{|\mathcal{U}|}\}$ denote the set of users and $\mathcal{V}=\{v_1,v_2,\dots,v_{|\mathcal{V}|}\}$ denote the item vocabulary. For each user $u\in\mathcal{U}$, we represent the historical interaction sequence as $\mathcal{H}_u=(v_1,v_2,\dots,v_T)$, where items are ordered chronologically. The task is to predict the next item $v_{T+1}$ conditioned on $\mathcal{H}_u$. We use a T5 encoder-decoder backbone~\cite{raffel2020exploring}: the encoder maps $\mathcal{H}_u$ into contextual hidden states, from which we denote the encoder-side user context vector as $\mathbf{q}_u$, and the decoder autoregressively generates the semantic ID tokens.

Following semantic ID-based generative recommendation, each item $v\in\mathcal{V}$ is represented by a hierarchical semantic ID sequence $S_v=(c_1,c_2,\dots,c_L)$, where $L$ is the semantic ID length and $c_i\in\{1,\dots,K\}$ denotes the discrete code index at the $i$-th position. The semantic IDs are produced by RQ-VAE through residual quantization. Specifically, each position is associated with a codebook $\mathcal{C}_i=\{e_{i,1},e_{i,2},\dots,e_{i,K}\}$, where $e_{i,k}$ is the codeword representation corresponding to code index $k$ at position $i$. The collection of these codeword representations defines the semantic ID representation space for item generation.

Generative recommendation predicts the semantic ID sequence of the next item autoregressively. For a target item with semantic ID sequence $S=(c_1,c_2,\dots,c_L)$, the generation probability conditioned on the user context is factorized as
\begin{equation}
\label{eq:generation_prob}
P(S\mid\mathbf{q}_u)=\prod_{i=1}^{L}P(c_i\mid c_{<i},\mathbf{q}_u),
\end{equation}
where $c_{<i}=(c_1,c_2,\dots,c_{i-1})$ denotes the previously generated semantic ID prefix. The standard training objective is the autoregressive cross-entropy loss:
\begin{equation}
\label{eq:ce_loss}
\mathcal{L}_{\text{CE}}
=
-\sum_{i=1}^{L}\log P(c_i\mid c_{<i},\mathbf{q}_u).
\end{equation}
During decoding, the generated prefix is constrained by a generation trie built from valid item semantic IDs, ensuring that each generated sequence corresponds to an item in the catalog.

\subsection{Latent Reasoning for Semantic ID Generation}
Standard semantic ID-based generative recommendation directly predicts each semantic ID token from the T5 decoder hidden state. This paradigm is efficient, but the computation used for each token prediction is limited by the fixed depth of the decoder. To exploit test-time scaling without generating explicit reasoning tokens, latent reasoning performs additional hidden-state computation before committing a semantic ID token.

Given the user context $\mathbf{q}_u$ and the committed prefix $c_{<i}$, the T5 decoder first produces an initial representation for predicting the $i$-th semantic ID token:
\begin{equation}
{H}^{(0)}_i = f_{\theta}(\mathbf{q}_u,c_{<i}).
\end{equation}
A latent reasoning mechanism then refines this representation for $k_i$ steps:
\begin{equation}
{H}^{(j)}_i = R_{\phi}^{(j)}({H}^{(j-1)}_i,\mathbf{q}_u,c_{<i}), 
\quad j=1,\dots,k_i,
\end{equation}
and predicts the current semantic ID token from the final refined representation:
\begin{equation}
P(c_i\mid c_{<i},\mathbf{q}_u)
=
\operatorname{softmax}\left(g_i({H}^{(k_i)}_i)\right).
\end{equation}
Here, $f_{\theta}$ denotes the T5 decoder that maps the user context and committed prefix to an initial hidden representation; $R_{\phi}^{(j)}$ denotes the $j$-th latent refinement operation; and $g_i(\cdot)$ denotes the prediction head that maps the refined representation to logits over the $i$-th semantic ID codebook.

Unlike explicit reasoning-token generation, the intermediate latent states are not appended to the generated sequence. They are used privately within the current semantic ID position, and only the committed token $c_i$ is exposed to subsequent positions. Under this formulation, two questions become central: how to design the refinement operation $R_{\phi}^{(j)}$ for multi-step latent refinement in a stable way, and how to determine how many refinement steps $k_i$ should be used at each semantic ID position according to its information contribution.

\section{The Proposed Method}
\label{sec:method}

In this section, we present IBA, an Information-Gain Budget Allocation framework for semantic ID-based generative recommendation. We first provide the motivation analysis, and then introduce the framework and its training procedure.

\subsection{Motivation Analysis}
\label{subsec:ig_analysis}

To motivate IBA, we first analyze the structural information contribution of different semantic ID positions. This analysis is model-independent and estimated from the training data: it does not determine the exact number of refinement steps for each user sequence, but provides a position-wise IG prior for the budget allocation module introduced later. Following previous work~\cite{lin2025igd}, we estimate information-gain offline from the training set only, without relying on model predictions. Specifically, for each item $v_i$, we use its empirical frequency in the training interactions as the prior probability $p(v_i)=n_{v_i}/C$, where $n_{v_i}$ is the number of interactions involving item $v_i$ and $C$ is the total number of interactions. For each semantic ID position $t$, we define the candidate set $\mathcal{S}(y_{1:t})$ as all training items whose semantic ID sequences share the prefix $y_{1:t}$. The conditional entropy of the candidate set given the prefix is computed as
\begin{equation}
\label{eq:conditional_entropy}
\mathcal{E}(y_{1:t}) = -\sum_{v_i \in \mathcal{S}(y_{1:t})} p_t(v_i) \log_2 p_t(v_i),
\end{equation}
where $p_t(v_i)=p(v_i)/\sum_{v_j \in \mathcal{S}(y_{1:t})}p(v_j)$ is the normalized prior within the candidate set. The information-gain at position $t$ is then defined as the entropy reduction:
\begin{equation}
\label{eq:ig_definition}
IG_t = \mathcal{E}(y_{1:t-1}) - \mathcal{E}(y_{1:t}).
\end{equation}
This metric measures how much the uncertainty over the candidate item set is reduced after observing the token at position $t$, and therefore quantifies the contribution of that position to identifying the target item. Since $IG_t$ is computed from training interactions and semantic-ID prefixes rather than model outputs, it characterizes the global discriminative structure of semantic ID generation.

\begin{figure}[!t]
\centering
\includegraphics[width=\columnwidth]{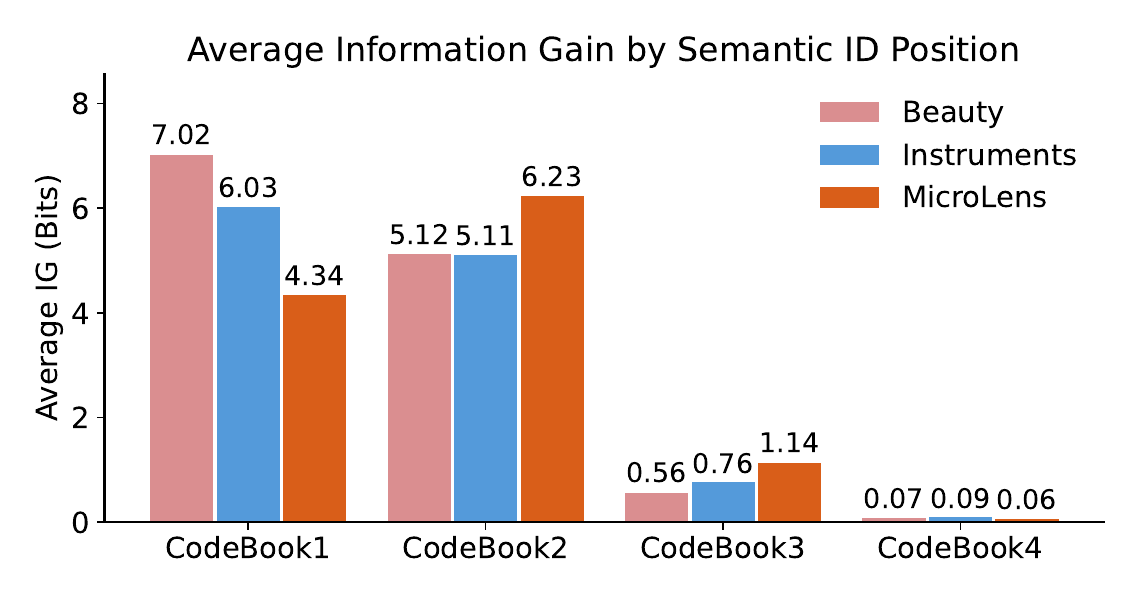}
\caption{Mean information-gain (IG) at each semantic ID position, estimated from the training set across three datasets.}
\label{fig:ig_bar}
\end{figure}

We aggregate $IG_t$ across items by position and report the mean values in Fig.~\ref{fig:ig_bar}. Instruments and Beauty show a monotonic decrease from position 1 to position 4, while MicroLens has a slightly higher mean IG at position 2 than position 1 but still shows a clear decline at later positions. These results indicate that semantic ID tokens at early positions have higher IG, which suggests that earlier tokens reduce more uncertainty about the target item.

\begin{figure*}[!t]
\centering
\includegraphics[width=0.9\textwidth]{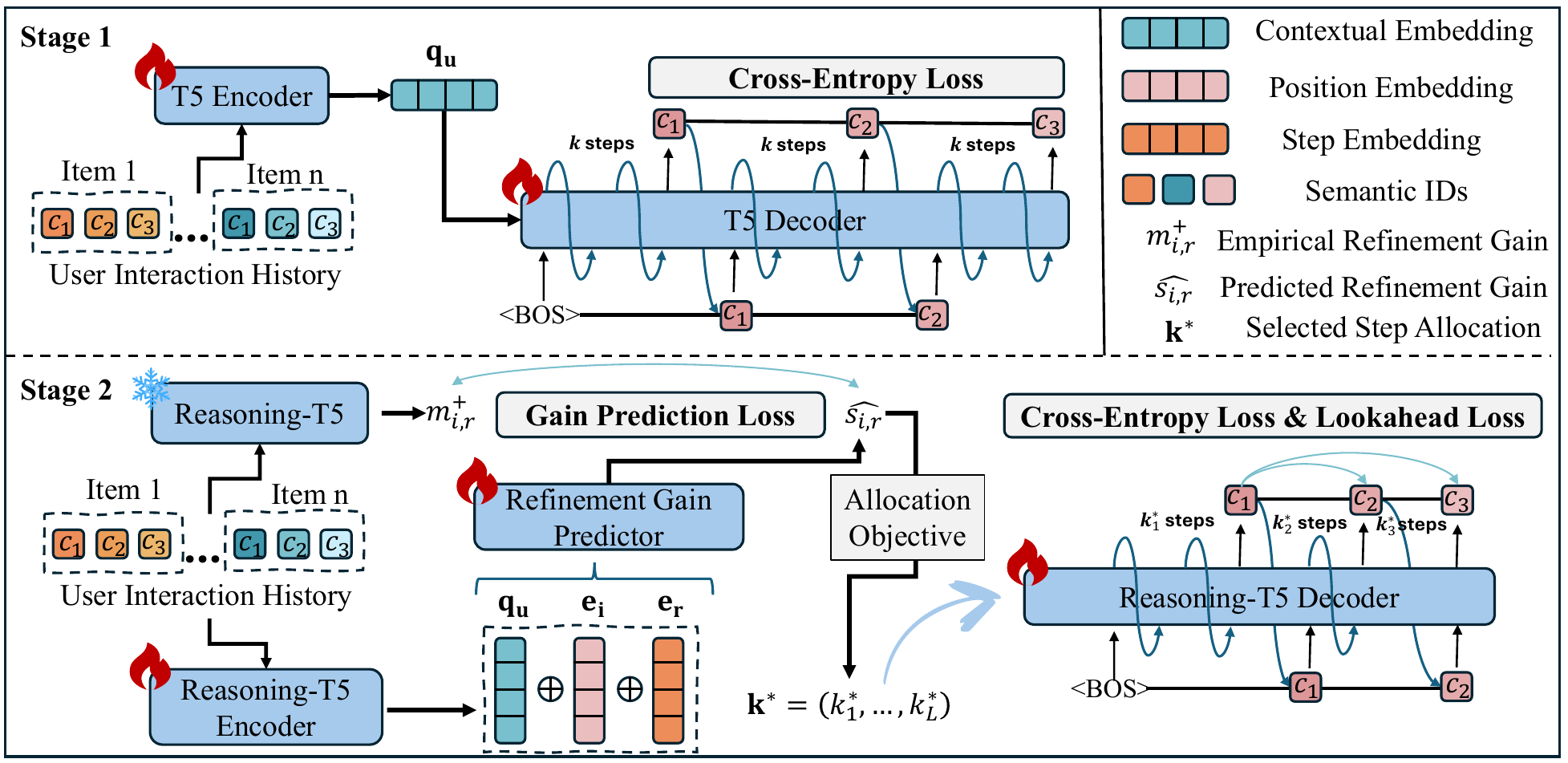}
\caption{Overview of the IBA framework. T5 denotes the original encoder-decoder backbone, while Reasoning-T5 denotes the Stage~1 model trained with a uniform refinement depth and reused in Stage~2.}
\label{fig:framework}
\end{figure*}

We next clarify why positions with higher IG are more valuable targets for additional latent refinement. Let $q_t(k)$ denote the probability of correctly predicting the semantic ID token at position $t$ after using $k$ refinement steps. We define the expected refinement utility at position $t$ as
\begin{equation}
\label{eq:refinement_utility}
U_t(k)=q_t(k)IG_t.
\end{equation}
This utility measures the expected contribution of position $t$ to correctly identifying the target item: $q_t(k)$ captures the probability that the token is predicted correctly, while $IG_t$ captures how much item-level uncertainty is reduced once this prediction is correct. The utility of one additional refinement step is therefore
\begin{equation}
\label{eq:marginal_refinement_utility}
\begin{aligned}
\Delta U_t(k)
&=
U_t(k+1)-U_t(k) \\
&=
\big(q_t(k+1)-q_t(k)\big)IG_t
=
\rho_t(k)IG_t,
\end{aligned}
\end{equation}
where $\rho_t(k)$ denotes the token-level improvement brought by the additional refinement step. This expression shows that a refinement step is more beneficial when a position has higher information-gain and the additional refinement step at this position improves token-level accuracy more. Within a moderate range of refinement steps, our step-wise refinement analysis in Section~\ref{subsubsec:refinement_dynamics} later verifies this pattern: early semantic ID positions improve token-level accuracy more from additional refinement, whereas later positions are already close to saturation and yield negligible further improvements. Therefore, earlier positions tend to offer larger expected refinement benefits, motivating IBA to allocate more refinement steps to early positions. Since different user sequences may still require slightly different budget allocation, IBA learns the allocation for each user sequence under this prior. We next describe how IBA implements this idea.

\subsection{Overview of IBA}
\label{subsec:iba_overview}

\begin{figure}[!t]
\centering
\includegraphics[width=3.5in]{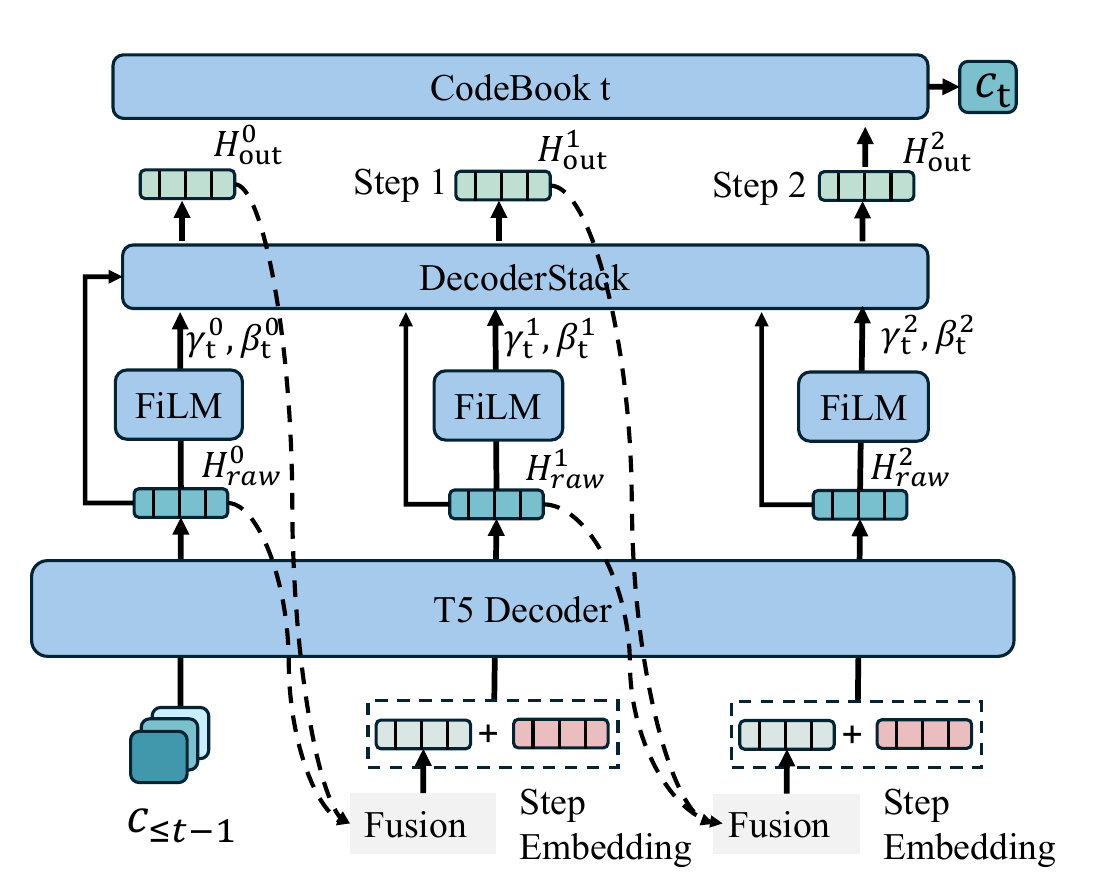}
\caption{Architecture of the Dual-Axis Refinement Module.}
\label{fig:correction_module}
\end{figure}

IBA is built on top of the T5 encoder-decoder backbone and allocates latent refinement steps across semantic ID positions. Before introducing the step allocation strategy, we first introduce a Dual-Axis Refinement Module to keep the reasoning process stable when different positions receive different numbers of refinement steps. The detailed architecture of this module is shown in Fig.~\ref{fig:correction_module}.

Based on the observed position-wise IG pattern, IBA then uses a Budget Allocation Module to select the step allocation with the largest expected gain for each user sequence under a fixed total budget. Finally, to train the introduced modules, IBA adopts a two-stage training strategy that first obtains empirical gain signals and then optimizes the model with the selected allocation. The following subsections describe these components in detail. The complete framework is shown in Fig.~\ref{fig:framework}.

\subsection{Dual-Axis Refinement}
\label{subsec:dual_axis_refinement}

The information-gain pattern suggests that different semantic ID positions should receive different numbers of refinement steps. To make this refinement pattern stable, the Dual-Axis Refinement Module operates along two complementary axes: horizontal multi-step reasoning progressively updates the hidden state within a position, while vertical semantic alignment keeps the intermediate state at each step close to the semantic ID space of that position. Given the user context $\mathbf{q}_u$ and the committed prefix, the T5 decoder first produces an initial hidden representation for the current semantic ID position. The module then updates this representation through the two components described below, and the number of refinement steps is provided by the step allocation strategy in Section~\ref{subsec:step_allocation}.

\subsubsection{Vertical Semantic Alignment Refinement}
\label{subsubsec:vertical_semantic_alignment_refinement}

Instead of directly projecting the last-layer decoder hidden state to semantic ID logits, we vertically align the intermediate representation at every refinement step with the semantic ID representation space. This vertical operation keeps the hidden state aligned with the position-specific semantic space at each refinement step, reducing semantic drift during horizontal reasoning. We further analyze this representation gap and the role of vertical alignment in Section~\ref{subsubsec:visualization}. Specifically, at position $t$, the model performs latent refinement over steps indexed by $j$. Let $H_{\text{raw},t}^{(j)}$ denote the raw decoder hidden state at refinement step $j$, where $j=0$ corresponds to the initial decoder representation produced from $\mathbf{q}_u$ and the committed prefix. The prediction target at this position is a discrete code index $c_t$ whose representation comes from the codebook $\mathcal{C}_t$ for position $t$.

At each refinement step $j$, we pass the raw decoder hidden state through an alignment stack built with the T5 decoder-layer architecture:
\begin{equation}
\label{eq:semantic_alignment}
H_{\text{align},t}^{(j)}
=
\mathrm{DecoderStack}_{\text{align}}\big(H_{\text{raw},t}^{(j)}\big),
\end{equation}
where $H_{\text{align},t}^{(j)}$ denotes the semantically aligned representation, and $\mathrm{DecoderStack}_{\text{align}}$ denotes a stack that reuses the layer architecture of the T5 decoder. This design preserves the contextual modeling structure of the backbone while providing additional hidden-state computation specialized for semantic alignment.

To allow different refinement steps to apply different alignment intensities, we further use a Feature-wise Linear Modulation (FiLM) mechanism~\cite{perez2018film}. For each step $j$, a step-specific MLP produces scale and shift parameters conditioned on the raw hidden state:
\begin{equation}
\label{eq:film_params}
\boldsymbol{\gamma}_{t}^{(j)}, \boldsymbol{\beta}_{t}^{(j)}
=
\mathrm{MLP}^{(j)}\big(H_{\text{raw},t}^{(j)}\big),
\end{equation}
where $\boldsymbol{\gamma}_{t}^{(j)}$ and $\boldsymbol{\beta}_{t}^{(j)}$ are $d$-dimensional modulation vectors. The final aligned representation is computed as
\begin{equation}
\label{eq:film_modulation}
H_{\text{out},t}^{(j)}
=
\big(\mathbf{1}+\boldsymbol{\gamma}_{t}^{(j)}\big)\odot H_{\text{align},t}^{(j)}
+
\boldsymbol{\beta}_{t}^{(j)},
\end{equation}
where $\odot$ denotes element-wise multiplication. The residual scaling form $(\mathbf{1}+\boldsymbol{\gamma}_{t}^{(j)})$ preserves the aligned signal when $\boldsymbol{\gamma}_{t}^{(j)}=\mathbf{0}$, while allowing the model to amplify or suppress specific semantic dimensions according to the current refinement step. The resulting representation $H_{\text{out},t}^{(j)}$ is used for semantic ID prediction and also serves as the refined signal for the next step update when additional refinement step is required.

\subsubsection{Horizontal Hidden Refinement}
\label{subsubsec:horizontal_refinement}

The horizontal hidden refinement component performs iterative latent reasoning within the same semantic ID position. Inspired by residual connections~\cite{he2016deep}, we avoid feeding only the previous aligned representation into the next refinement step. A direct recurrence of this form may accumulate errors, especially when the refinement trajectory drifts away from the target semantic ID representation space. To stabilize the recurrence, we retain the raw decoder output from the previous step, $H_{\text{raw},t}^{(j-1)}$, as an anchor and combine it with the previous aligned representation. For refinement step $j\in\{1,\dots,k_t^{*}\}$ at position $t$, the input representation is constructed as
\begin{equation}
\label{eq:horizontal_input}
H_{\text{in},t}^{(j)}
=
(1-\gamma_{\mathrm{rec}})\cdot H_{\text{raw},t}^{(j-1)}
+
\gamma_{\mathrm{rec}}\cdot H_{\text{out},t}^{(j-1)}
+
\mathbf{a}_j,
\end{equation}
where $k_t^{*}$ is the number of refinement steps selected for position $t$, $\gamma_{\mathrm{rec}}$ is the recurrence fusion coefficient, and $\mathbf{a}_j\in\mathbb{R}^d$ is a learnable step embedding that identifies the $j$-th refinement step.

The fused representation is processed by the T5 decoder to produce the next raw hidden state:
\begin{equation}
\label{eq:horizontal_decoder}
H_{\text{raw},t}^{(j)}
=
\mathrm{Decoder}\big(H_{\text{in},t}^{(j)}\big).
\end{equation}
This raw hidden state is again passed through the vertical semantic alignment component in Eqs.~\eqref{eq:semantic_alignment}--\eqref{eq:film_modulation}, yielding $H_{\text{out},t}^{(j)}$. After $k_t^{*}$ refinement steps, the final representation is projected to semantic ID logits:
\begin{equation}
\label{eq:final_logits}
\mathbf{z}_t
=
\mathrm{LMHead}_{t}\big(H_{\text{out},t}^{(k_t^{*})}\big).
\end{equation}

When $k_t^{*}=0$, the model directly uses the initial refined representation $H_{\text{out},t}^{(0)}$ for prediction. Thus, the same Dual-Axis Refinement Module supports positions assigned different numbers of refinement steps under a unified formulation in a stable way.

\subsection{Budget Allocation}
\label{subsec:step_allocation}

This module determines how many latent refinement steps to assign to each semantic ID position. We refer to the total number of available refinement steps as the budget. Formally, $k_i$ denotes the number of refinement steps assigned to position $i$, and $B$ denotes the total step budget across the whole semantic ID sequence. Guided by the information-gain analysis in Section~\ref{subsec:ig_analysis}, IBA uses a prior that assigns more refinement steps to early positions while allowing the final allocation to vary across user sequences.

For a user sequence $u$, IBA uses the encoder-side user context vector $\mathbf{q}_u$ defined in Section~\ref{subsec:preliminaries}. Before decoding the target semantic ID, IBA evaluates the refinement gain by a lightweight MLP predictor for each candidate refinement step $r\in\{1,\dots,K_{\max}\}$ at each semantic ID position $i$:
\begin{equation}
\label{eq:budget_predictor}
\widehat{s}_{i,r}
=
\mathrm{Softplus}
\left(
g_{\psi}
\left(
\left[
\mathbf{q}_u;\mathbf{e}_i;\mathbf{e}_r
\right]
\right)
\right),
\end{equation}
where $g_{\psi}$ denotes the refinement gain predictor, and $\mathbf{e}_i$ and $\mathbf{e}_r$ are learnable embeddings for the semantic ID position and the refinement step, respectively. The predicted score $\widehat{s}_{i,r}$ estimates the user-conditioned gain score of assigning the $r$-th refinement step to position $i$ for the current user sequence. The Softplus activation keeps this score non-negative.

To make this score meaningful, the refinement gain predictor is supervised by empirical refinement gains estimated from the Stage~1 model. The overall two-stage training procedure is described in Section~\ref{subsec:two_stage_training}. After training with the same refinement steps across positions, we unroll the Stage~1 checkpoint up to $K_{\max}$ candidate refinement steps and compute the prediction loss at each step. Let $\mathbf{z}_i^{(r)}$ be the logits produced for position $i$ after $r$ refinement steps, and define
\begin{equation}
\label{eq:refinement_loss}
\ell_i^{(r)}
=
\mathcal{L}_{\mathrm{CE}}
\left(
\mathbf{z}_i^{(r)}, c_i
\right).
\end{equation}
Following the refinement utility analysis in Section~\ref{subsec:ig_analysis}, the gain of one additional refinement step is defined as $\Delta U_t(k)=\rho_t(k)IG_t$, where $\rho_t(k)$ denotes the token-level accuracy improvement. In the refinement gain predictor, we replace this accuracy improvement term with the reduction in token-level prediction loss. This loss reduction also reflects how much the additional refinement step improves token prediction, and it provides a continuous training signal that is easier to compute for supervising the predictor. The clipped empirical refinement gain of the $r$-th refinement step is
\begin{equation}
\label{eq:marginal_gain}
m_{i,r}^{+}
=
\max
\left(
0,\,
\ell_i^{(r-1)}-\ell_i^{(r)}
\right)
\cdot
IG_i,
\end{equation}
which measures the utility brought by the additional refinement step at position $i$. We train the predictor with
\begin{equation}
\label{eq:gain_prediction_loss}
\mathcal{L}_{\mathrm{gain}}
=
\sum_{i=1}^{L}
\sum_{r=1}^{K_{\max}}
\left(
\widehat{s}_{i,r}
-
\mathrm{sg}
\left(
m_{i,r}^{+}
\right)
\right)^2,
\end{equation}
where $\mathrm{sg}(\cdot)$ denotes stop-gradient. This loss calibrates the predicted score $\widehat{s}_{i,r}$ to refinement gain from the Stage~1 model, so a larger score indicates that the corresponding refinement step is expected to be more useful for that position and user sequence.

Given these predicted refinement gain scores, IBA selects the step allocation by solving an allocation objective with a prior penalty:
\begin{equation}
\label{eq:budget_objective}
\mathbf{k}^{*}
=
\arg\max_{\mathbf{k}\in\mathcal{K}}
\left[
\sum_{i=1}^{L}
\sum_{r=1}^{k_i}
\widehat{s}_{i,r}
-
\lambda_{\mathrm{P}}
\left\|
\mathbf{k}
-
\mathbf{k}^{\mathrm{prior}}
\right\|_{1}
\right],
\end{equation}
where $\mathbf{k}^{*}=(k_1^{*},\dots,k_L^{*})$ is the selected step allocation and $k_i^{*}$ denotes the number of refinement steps assigned to the $i$-th semantic ID position. The first term accumulates the predicted refinement gains, while the second term softly biases the selected allocation toward a prior step allocation $\mathbf{k}^{\mathrm{prior}}$. The coefficient $\lambda_{\mathrm{P}}$ controls the strength of this prior, so the selected allocation may still deviate from $\mathbf{k}^{\mathrm{prior}}$ when the predicted gains support another allocation. The feasible set is
\begin{equation}
\label{eq:budget_constraint}
\mathcal{K}
=
\left\{
\mathbf{k}
\mid
0\leq k_i\leq K_{\max},\;
\sum_{i=1}^{L}k_i=B
\right\},
\end{equation}
where $B$ is the total number of refinement steps available for one semantic ID sequence. The constraint $\sum_i k_i=B$ fixes the total number of latent refinement steps, while the prior term in Eq.~\eqref{eq:budget_objective} softly encourages the selected allocation to follow the overall information-gain pattern. Because $L$ and $K_{\max}$ are small in semantic ID generation, Eq.~\eqref{eq:budget_objective} can be solved by enumerating all feasible allocations with negligible overhead. The selected step allocation is computed once for each user sequence and shared across beam-search candidates, keeping the decoding process deterministic and efficient. In Section~\ref{subsec:budget_vs_fixed}, we further examine different choices of $\mathbf{k}^{\mathrm{prior}}$, including a uniform prior, a prior that assigns more steps to later positions, and a prior that assigns more steps to early positions.

\subsection{Two-Stage Training}
\label{subsec:two_stage_training}

IBA uses a two-stage training protocol to make the refinement operation and step allocation learnable in sequence. Stage~1 obtains a uniformly refined checkpoint and empirical gain signals, while Stage~2 trains the refinement gain predictor and fine-tunes the recommender under the selected step allocation and the lookahead objective.

\subsubsection{Stage~1 Training}
\label{subsubsec:uniform_step_pretraining}

During Stage~1 training, every semantic ID position is assigned the same number of refinement steps. Formally, for all positions $i\in\{1,\dots,L\}$, we set
\begin{equation}
\label{eq:uniform_steps}
k_i = K,
\end{equation}
where $K$ is a small constant, such as $K=2$. This setting gives all positions the same refinement process and avoids introducing different refinement steps across positions before the model has learned the basic refinement operation. The model is trained with standard cross-entropy loss on semantic ID prediction under this refinement process.

\subsubsection{Stage~2 Training}
\label{subsubsec:stairwise_finetuning}

During Stage~2 training, we initialize the model from the Stage~1 checkpoint and introduce the refinement gain predictor defined in Eq.~\eqref{eq:budget_predictor}. For each user sequence, the predictor produces gain scores $\widehat{s}_{i,r}$ for candidate refinement steps, and the allocation objective in Eq.~\eqref{eq:budget_objective} selects $\mathbf{k}^{*}=(k_1^{*},\dots,k_L^{*})$ to maximize the total predicted gain under the budget constraint. This stage optimizes gain prediction with $\mathcal{L}_{\mathrm{gain}}$ and fine-tunes the recommender under the same selected step allocation used during inference.

Given the selected step allocation, Stage~2 optimizes the terminal prediction associated with the selected number of refinement steps at each position. Let $\mathbf{z}_i^{(k_i^{*})}$ denote the logits produced at position $i$ after $k_i^{*}$ refinement steps. The main loss is
\begin{equation}
\label{eq:stage2_main_loss}
\mathcal{L}_{\text{main}}
=
\frac{1}{L}
\sum_{i=1}^{L}
\mathcal{L}_{\text{CE}}\big(\mathbf{z}_i^{(k_i^{*})}, c_i\big).
\end{equation}

IBA usually allocates more refinement steps to earlier semantic ID positions, which receive more computation and contain much of the information needed to identify the target item. Their refined representations should therefore support not only the current token prediction but also later semantic ID predictions. We introduce a prefix lookahead objective as an auxiliary regularizer, encouraging early refined representations to remain useful for predicting nearby future semantic ID tokens. For each valid lookahead offset $\delta\in\{1,\dots,\Delta\}$, we add a learnable offset vector $\mathbf{o}_\delta$ to the final refined representation at position $t$ and use the LM head to predict the future token $c_{t+\delta}$:
\begin{equation}
\label{eq:lookahead_prob}
P_\delta(c_{t+\delta}\mid \mathbf{q}_u, c_{<t})
=
\mathrm{softmax}
\left(
\mathrm{LMHead}_{t+\delta}\big(H_{\text{out},t}^{(k_t^{*})} + \mathbf{o}_\delta\big)
\right),
\end{equation}
where $\mathrm{LMHead}_{t+\delta}(\cdot)$ denotes the shared prediction head for the semantic ID codebook at position $t+\delta$, and $\Delta$ is the maximum lookahead offset. The lookahead loss is computed over all valid position-offset pairs:
\begin{equation}
\label{eq:lookahead_loss}
\mathcal{L}_{\mathrm{LA}}
=
\frac{1}{|\Omega|}
\sum_{(t,\delta)\in\Omega}
\mathcal{L}_{\mathrm{CE}}
\big(
P_\delta(c_{t+\delta}\mid \mathbf{q}_u, c_{<t}),
c_{t+\delta}
\big),
\end{equation}
where
\begin{equation}
\label{eq:lookahead_pairs}
\Omega
=
\{(t,\delta)\mid 1\leq t\leq L,\;1\leq\delta\leq\Delta,\;t+\delta\leq L\}.
\end{equation}
The full Stage~2 objective is
\begin{equation}
\label{eq:stage2_loss}
\mathcal{L}_{\text{stage2}}
=
\mathcal{L}_{\text{main}}
+
\lambda_{\mathrm{LA}}\mathcal{L}_{\mathrm{LA}}
+
\lambda_{\mathrm{gain}}\mathcal{L}_{\mathrm{gain}},
\end{equation}
where $\lambda_{\mathrm{LA}}$ and $\lambda_{\mathrm{gain}}$ control the contributions of the lookahead objective and gain prediction loss, respectively.

\section{Experiments}
\label{sec:experiments}

In this section, we conduct experiments to evaluate the effectiveness of IBA and analyze how its learned step allocation affects both accuracy and efficiency. We aim to answer the following research questions:

\begin{list}{}{\setlength{\leftmargin}{3.2em}
\setlength{\labelwidth}{2.6em}
\setlength{\labelsep}{0.6em}
\setlength{\itemsep}{1pt}
\setlength{\parsep}{0pt}
\setlength{\topsep}{2pt}}
\item[\textbf{RQ1:}] How does IBA compare with existing state-of-the-art recommendation methods?
\item[\textbf{RQ2:}] What are the individual contributions of semantic alignment refinement, the refinement gain predictor, and the lookahead auxiliary objective?
\item[\textbf{RQ3:}] How sensitive is IBA to key hyperparameters, including the lookahead weight $\lambda_{\mathrm{LA}}$ and the recurrence fusion coefficient $\gamma_{\mathrm{rec}}$?
\item[\textbf{RQ4:}] How does semantic alignment refinement affect the hidden space, and how much token-level hit-rate gain does each position obtain from refinement steps?
\end{list}

\subsection{Experimental Setup}
\label{subsec:experimental_setup}

\subsubsection{Datasets}
\label{subsubsec:datasets}

We conduct experiments on three publicly available datasets spanning different domains: Instruments, Beauty, and MicroLens. The Instruments and Beauty datasets are derived from the Amazon review corpus~\cite{ni2019justifying}. Instruments contains user interactions with musical instruments, and Beauty contains user interactions with beauty products. MicroLens is a micro-video recommendation dataset that contains user interactions with micro-videos~\cite{ni2025content}. Following the standard protocol in generative recommendation~\cite{rajput2023recommender, wang2024learnable}, we apply a leave-one-out splitting strategy for each dataset. We filter out users with fewer than 5 interactions. During training, we limit the number of items in a user's history to 20. Item semantic IDs are generated via residual vector quantization (RQ-VAE)~\cite{lee2022autoregressive}, with each item represented as a 4-token hierarchical sequence. Table~\ref{tab:dataset_stats} summarizes the statistics of the three datasets after preprocessing.

\begin{table}[!t]
\renewcommand{\arraystretch}{1.3}
\caption{Dataset Statistics}
\label{tab:dataset_stats}
\centering
\begin{tabular}{ccccc}
\toprule
\textbf{Dataset} & \textbf{\#Users} & \textbf{\#Items} & \textbf{\#Interactions} & \textbf{\#Avg.n} \\
\midrule
Instruments & 24,772 & 9,922 & 206,153 & 8.32\\
Beauty & 22,363 & 12,101 & 198,502 & 8.88  \\
MicroLens & 13,210 & 5,312 & 78,255 & 5.92\\
\bottomrule
\end{tabular}
\end{table}

\subsubsection{Baselines}
\label{subsubsec:baselines}

We compare IBA with four groups of baselines: traditional sequential recommenders, including SASRec~\cite{kang2018self} and CASER~\cite{tang2018personalized}; LLM-based recommendation with latent reasoning, represented by LatentR\textsuperscript{3}~\cite{zhang2025reinforced}; semantic ID-based generative recommenders, including TIGER~\cite{rajput2023recommender} and LETTER~\cite{wang2024learnable}; and semantic ID-based generative recommendation with latent reasoning, represented by CARE~\cite{lin2026bringing}. To evaluate IBA as a modular enhancement, we apply it on top of TIGER and LETTER. The ``+'' notation in our tables indicates that IBA is added to the corresponding base model.

\subsubsection{Evaluation Metrics}
\label{subsubsec:metrics}

Following standard practices in generative recommendation~\cite{rajput2023recommender, zhou2025onerec}, we evaluate all methods using Recall and Normalized Discounted Cumulative Gain (NDCG) at cutoffs of 5 and 10, denoted as R@5, R@10, N@5, and N@10, respectively. Since the leave-one-out protocol yields exactly one ground-truth item per user, the ideal DCG is always 1.0, making NDCG equivalent to the discounted hit position. Higher values for all metrics indicate better recommendation performance.

\subsubsection{Implementation Details}
\label{subsubsec:implementation_details}

We instantiate the backbone with T5-small~\cite{raffel2020exploring} in all experiments. Item semantic IDs are produced by an RQ-VAE tokenizer, where each item's text information (e.g., title, category, and description) is encoded into a dense vector using Sentence-T5~\cite{ni2022sentence}. The model is trained using a two-stage protocol. In Stage 1, we train the model with uniform refinement steps ($K=2$) for 200 epochs, using a learning rate of $5 \times 10^{-4}$, batch size of 128, and gradient accumulation steps of 2. In Stage 2, we fine-tune the Stage 1 checkpoint with the learned refinement gain predictor for 50 epochs, using a lower learning rate of $5 \times 10^{-5}$. Unless otherwise specified, the prior step allocation is set to $\mathbf{k}^{\mathrm{prior}}=(3,2,1,0)$, corresponding to a total step budget of 6, and the prior weight is set to $\lambda_{\mathrm{P}}=0.1$. The lookahead auxiliary objective uses a maximum offset of $\Delta = 2$ with weight $\lambda_{\mathrm{LA}} = 0.15$. The recurrence fusion coefficient is set to $\gamma_{\mathrm{rec}} = 0.5$. The semantic alignment stack uses $L_{\text{align}}=5$ layers and follows the architecture of the T5 decoder blocks.

During inference, we employ beam search with 20 beams and a maximum of 10 generated tokens. The generation is constrained by a prefix trie to ensure that all generated semantic ID sequences correspond to valid items in the catalog. All models are trained and evaluated on an NVIDIA RTX 4090 GPU (24 GB) with a random seed of 42 for reproducibility.

\subsection{Main Results}
\label{subsec:main_results}

Table~\ref{tab:main_results} summarizes the main results on three public datasets. From the table, we have the following observations:

\begin{table*}[!t]
\renewcommand{\arraystretch}{1.15}
\caption{Performance comparison of different methods on three public datasets. Best results in each column are boldfaced.}
\label{tab:main_results}
\centering
\begin{tabular}{l|cccc|cccc|cccc}
\hline
\hline
\multirow{2}{*}{\textbf{Method}} & \multicolumn{4}{c|}{\textbf{Instruments}} & \multicolumn{4}{c|}{\textbf{Beauty}} & \multicolumn{4}{c}{\textbf{MicroLens}} \\
\cline{2-13}
& \textbf{R@5} & \textbf{R@10} & \textbf{N@5} & \textbf{N@10} & \textbf{R@5} & \textbf{R@10} & \textbf{N@5} & \textbf{N@10} & \textbf{R@5} & \textbf{R@10} & \textbf{N@5} & \textbf{N@10} \\
\hline
SASRec & 0.0512 & 0.0746 & 0.0301 & 0.0373 & 0.0196 & 0.0368 & 0.0106 & 0.0160 & 0.0375 & 0.0698 & 0.0210 & 0.0314 \\
CASER & 0.0563 & 0.0746 & 0.0443 & 0.0502 & 0.0203 & 0.0340 & 0.0125 & 0.0170 & 0.0157 & 0.0301 & 0.0076 & 0.0122 \\
LatentR\textsuperscript{3} & 0.0749 & 0.0963 & 0.0641 & 0.0709 & 0.0268 & 0.0453 & 0.0161 & 0.0220 & 0.0388 & 0.0617 & 0.0248 & 0.0322 \\
CARE & 0.0572 & 0.0872 & 0.0413 & 0.0510 & 0.0229 & 0.0444 & 0.0135 & 0.0204 & 0.0372 & 0.0583 & 0.0233 & 0.0301 \\
\hline
TIGER & 0.0687 & 0.0853 & 0.0576 & 0.0629 & 0.0206 & 0.0355 & 0.0127 & 0.0174 & 0.0350 & 0.0561 & 0.0221 & 0.0289 \\
\rowcolor{starecgray}
\quad + IBA & \textbf{0.0780} & \textbf{0.0975} & \textbf{0.0663} & \textbf{0.0725} & \textbf{0.0277} & 0.0453 & \textbf{0.0173} & 0.0230 & 0.0431 & 0.0646 & 0.0278 & 0.0347 \\
\hline
LETTER & 0.0705 & 0.0882 & 0.0604 & 0.0661 & 0.0216 & 0.0362 & 0.0129 & 0.0176 & 0.0464 & 0.0729 & 0.0298 & 0.0382 \\
\rowcolor{starecgray}
\quad + IBA & 0.0766 & 0.0964 & 0.0656 & 0.0719 & 0.0274 & \textbf{0.0475} & 0.0172 & \textbf{0.0236} & \textbf{0.0540} & \textbf{0.0780} & \textbf{0.0360} & \textbf{0.0437} \\
\hline
\hline
\end{tabular}
\end{table*}

\begin{itemize}
\item[(1)] IBA consistently improves semantic ID-based generative recommenders. When applied to TIGER, IBA improves all metrics on all three datasets. The same pattern holds for LETTER: IBA improves Instruments, Beauty, and MicroLens across all reported metrics, showing that the proposed step allocation is not tied to a single base tokenizer or backbone.

\item[(2)] IBA achieves competitive or stronger performance than existing reasoning-enhanced baselines. Compared with CARE, both TIGER+IBA and LETTER+IBA obtain higher scores on every dataset and metric. Compared with LatentR\textsuperscript{3}, IBA-enhanced semantic ID models also achieve stronger overall results: TIGER+IBA slightly exceeds LatentR\textsuperscript{3} on Instruments and matches or improves it on Beauty, while LETTER+IBA gives the best overall results on Beauty R@10/N@10 and all MicroLens metrics.

\item[(3)] The strongest IBA-enhanced backbone differs across datasets. TIGER+IBA obtains the best results on Instruments and the best R@5/N@5 on Beauty, whereas LETTER+IBA performs best on Beauty R@10/N@10 and all MicroLens metrics.
\end{itemize}

\subsection{Effect of Prior Step Schedules}
\label{subsec:budget_vs_fixed}

Before analyzing model components, we first examine how the prior step schedule in Eq.~\eqref{eq:budget_objective} affects recommendation performance. The tuple $\mathbf{k}^{\mathrm{prior}}$ specifies the preferred number of refinement steps at each semantic ID position, and its sum corresponds to the total step budget used by the allocator. We include $\mathbf{k}^{\mathrm{prior}}=(2,2,2,2)$ as a uniform 8-step reference, where every position receives the same prior preference. We further compare two 6-step non-uniform priors: $(0,1,2,3)$ assigns more steps to later positions, while the default prior $(3,2,1,0)$ assigns more steps to early positions. This comparison tests whether performance depends only on using more refinement steps, or whether the direction of step allocation also matters.

Table~\ref{tab:budget_vs_fixed} summarizes the results. The uniform 8-step prior $(2,2,2,2)$ already provides a strong reference, but it is not the best choice. The prior $(0,1,2,3)$ performs worse on both datasets, especially on MicroLens, indicating that a non-uniform allocation can hurt performance when it conflicts with the position-wise IG pattern. In contrast, the prior $(3,2,1,0)$ achieves the best results on both datasets. These results suggest that allocating refinement steps according to the information structure is more effective than simply using a larger uniform step budget.

\begin{table}[!t]
\renewcommand{\arraystretch}{1.15}
\caption{Effect of different prior step schedules. Best results in each column are boldfaced.}
\label{tab:budget_vs_fixed}
\centering
\small
\begin{tabular}{lcccc}
\toprule
 & \multicolumn{2}{c}{\textbf{Instruments}} & \multicolumn{2}{c}{\textbf{MicroLens}} \\
\cmidrule(lr){2-3}\cmidrule(lr){4-5}
\textbf{Prior Step Schedule} & \textbf{R@10} & \textbf{N@10} & \textbf{R@10} & \textbf{N@10} \\
\midrule
IBA w/ $(2,2,2,2)$ & 0.0961 & 0.0718 & 0.0605 & 0.0335 \\
IBA w/ $(0,1,2,3)$ & 0.0947 & 0.0707 & 0.0597 & 0.0330 \\
IBA w/ $(3,2,1,0)$ & \textbf{0.0975} & \textbf{0.0725} & \textbf{0.0646} & \textbf{0.0347} \\
\bottomrule
\end{tabular}
\end{table}

\subsection{Ablation Studies}
\label{subsec:ablation_studies}

To answer RQ2, we conduct an ablation study on three key components in IBA: semantic alignment refinement, user-specific step allocation enabled by refinement gain prediction, and the lookahead auxiliary objective. We remove each component while keeping the rest of the framework unchanged, and report the results in Table~\ref{tab:ablation}.

\begin{table}[!t]
\renewcommand{\arraystretch}{1.15}
\caption{Ablation studies of IBA. Best results in each column are boldfaced.}
\label{tab:ablation}
\centering
\small
\begin{tabular}{lcccc}
\toprule
 & \multicolumn{2}{c}{\textbf{Instruments}} & \multicolumn{2}{c}{\textbf{MicroLens}} \\
\cmidrule(lr){2-3}\cmidrule(lr){4-5}
\textbf{Method} & \textbf{R@10} & \textbf{N@10} & \textbf{R@10} & \textbf{N@10} \\
\midrule
IBA (full) & \textbf{0.0975} & \textbf{0.0725} & \textbf{0.0646} & \textbf{0.0347} \\
\quad w/o Alignment & 0.0939 & 0.0724 & 0.0556 & 0.0303 \\
\quad w/o Predictor & 0.0958 & 0.0715 & 0.0628 & 0.0345 \\
\quad w/o Lookahead & 0.0948 & 0.0714 & 0.0612 & 0.0338 \\
\bottomrule
\end{tabular}
\end{table}

Table~\ref{tab:ablation} shows that the full model performs best on both datasets, and each removed component causes a performance drop.
\begin{itemize}
\item Removing semantic alignment refinement weakens the alignment between T5 decoder states and the semantic ID codeword space. This makes token prediction rely more directly on the raw decoder representation, which has a gap with semantic ID space.

\item Removing the predictor means that IBA no longer learns a step allocation for each user sequence. Instead, it uses the fixed allocation $(3,2,1,0)$ for all user sequences. This variant remains competitive, confirming that the information-gain prior provides a useful allocation direction, but it cannot adapt the refinement steps to the specific user context.

\item Removing the lookahead objective eliminates an auxiliary signal that encourages early refined representations to preserve information useful for later semantic ID positions. Without this signal, early-position refinement may focus too narrowly on the current token and provide weaker support for subsequent generation.
\end{itemize}

\subsection{Hyperparameter Sensitivity}
\label{subsec:hyperparameter_sensitivity}

To answer RQ3, we investigate the sensitivity of IBA to two key hyperparameters: the lookahead weight $\lambda_{\mathrm{LA}}$ and the recurrence fusion coefficient $\gamma_{\mathrm{rec}}$. All experiments are conducted on the Instruments and MicroLens datasets with the default configuration unless otherwise specified.

\subsubsection{Lookahead Weight $\lambda_{\mathrm{LA}}$}
\label{subsubsec:sensitivity_lookahead}

The lookahead weight $\lambda_{\mathrm{LA}}$ controls the contribution of the lookahead auxiliary loss relative to the cross-entropy loss. We vary $\lambda_{\mathrm{LA}}$ from 0 to 0.45 and report the performance in Fig.~\ref{fig:sensitivity_combined} (top row).

\subsubsection{Recurrence Fusion Coefficient $\gamma_{\mathrm{rec}}$}
\label{subsubsec:sensitivity_recurrence}

\begin{figure*}[!t]
\centering
\includegraphics[width=0.9\textwidth]{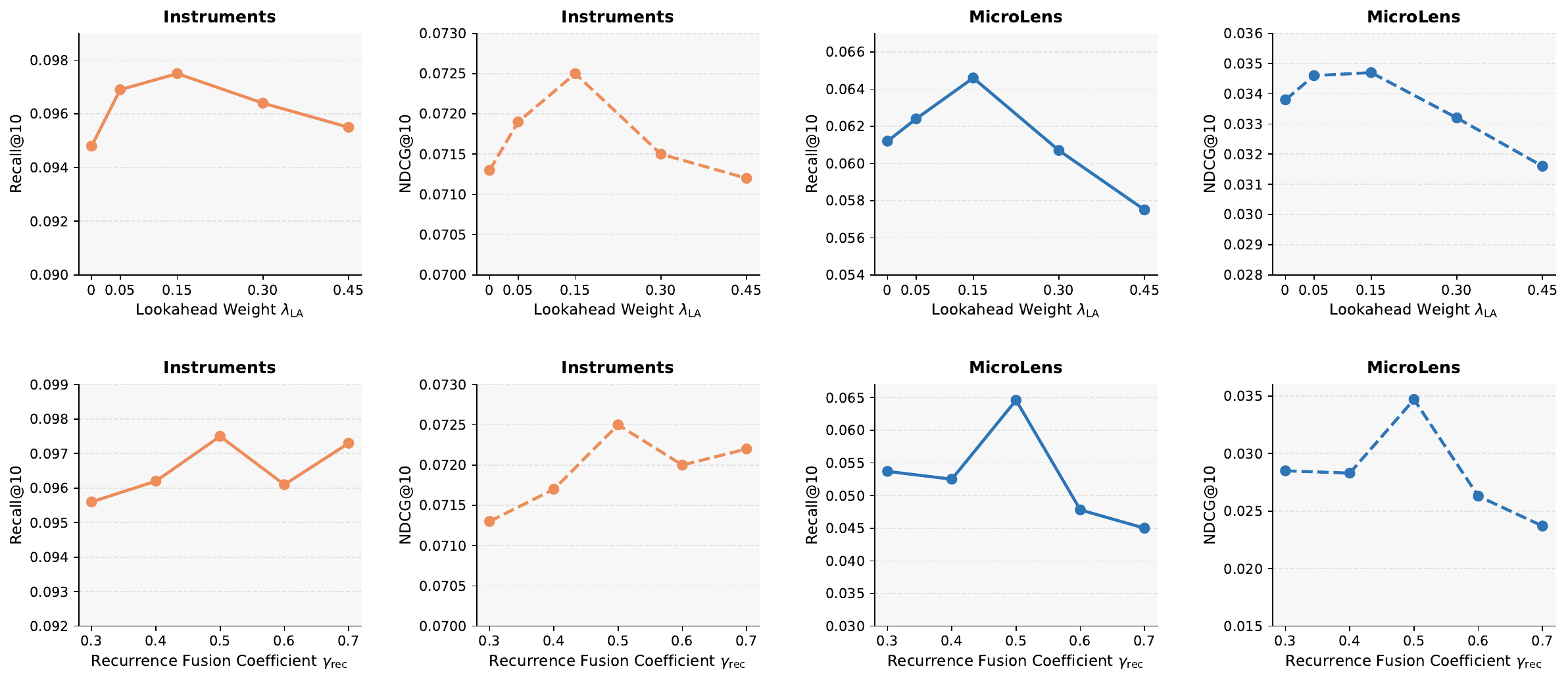}
\caption{Hyperparameter sensitivity analysis on Instruments and MicroLens. Top row: Lookahead weight $\lambda_{\mathrm{LA}}$; bottom row: Recurrence fusion coefficient $\gamma_{\mathrm{rec}}$. Each column shows a specific dataset and metric combination.}
\label{fig:sensitivity_combined}
\end{figure*}

The recurrence fusion coefficient $\gamma_{\mathrm{rec}}$ controls how much the previous refined representation is incorporated into the input of the current refinement step. We vary $\gamma_{\mathrm{rec}}$ from 0.3 to 0.7 and report the results in Fig.~\ref{fig:sensitivity_combined} (bottom row).

For the lookahead weight $\lambda_{\mathrm{LA}}$, Fig.~\ref{fig:sensitivity_combined} shows that removing the lookahead objective leads to weaker performance on both datasets. The recommendation performance improves as $\lambda_{\mathrm{LA}}$ increases from 0 to a moderate value, with the best overall results achieved at $\lambda_{\mathrm{LA}}=0.15$. However, further increasing the weight hurts performance, especially on MicroLens. These results indicate that the lookahead objective is useful as an auxiliary signal, but it should remain moderately weighted so that it does not dominate the main semantic ID prediction objective.

For the recurrence fusion coefficient $\gamma_{\mathrm{rec}}$, the best performance is also achieved at a moderate value, with $\gamma_{\mathrm{rec}}=0.5$ performing best on both datasets. Smaller values provide insufficient feedback from the previous refined representation, limiting iterative improvement. Larger values place too much weight on the refined state and can destabilize the refinement trajectory. The sensitivity is more pronounced on MicroLens, suggesting that the balance between the raw decoder state and the previous refined representation is important for stable latent refinement.

\subsection{Performance and Efficiency Analysis}
\label{subsec:efficiency_analysis}

We further evaluate the performance--efficiency trade-off of IBA on the Instruments dataset. Table~\ref{tab:efficiency} reports the parameter count, recommendation accuracy, and average inference time per test user under the same GPU, beam size, and decoding setting.

TIGER has the lowest latency because it directly predicts semantic ID tokens without additional latent refinement. Uniform-$k=2$ refinement and IBA both introduce the Dual-Axis Refinement Module and therefore have the same parameter count. The key difference is that uniform-$k=2$ uses eight refinement steps for each semantic ID sequence, whereas IBA uses a six-step allocation biased toward earlier positions. As shown in Table~\ref{tab:efficiency}, IBA reduces inference time compared with uniform refinement while maintaining slightly better recommendation accuracy. This result indicates that a step allocation consistent with the position-wise information-gain pattern is more effective: assigning more budget to earlier positions and reducing redundant refinement at later positions can lower the total computation and inference latency while also bringing a meaningful performance improvement.

\begin{table}[!t]
\renewcommand{\arraystretch}{1.15}
\caption{Performance and efficiency comparison on Instruments. Time/User denotes the average inference time per test user, measured over the model generation step on an NVIDIA RTX 4090. Best accuracy results are boldfaced.}
\label{tab:efficiency}
\centering
\small
\begin{tabular}{lcccc}
\toprule
\textbf{Method} & \textbf{Params} & \textbf{R@10} & \textbf{N@10} & \textbf{Time/User} \\
\midrule
TIGER & 60.9M & 0.0853 & 0.0629 & 20.99 ms \\
IBA w/ $(2,2,2,2)$ & 83.5M & 0.0961 & 0.0718 & 25.78 ms \\
\rowcolor{starecgray}
IBA w/ $(3,2,1,0)$ & 83.5M & \textbf{0.0975} & \textbf{0.0725} & 24.61 ms \\
\bottomrule
\end{tabular}
\end{table}

\subsection{Qualitative Analysis}
\label{subsec:qualitative_analysis}

To answer RQ4, we analyze both the geometric effect of semantic alignment refinement on the hidden space and the token-level hit-rate gain obtained by each semantic ID position across successive refinement steps. We first visualize the hidden representations before and after semantic alignment refinement, and then compare how Hit@5 changes from Step~0 to later refinement steps at each position.

\subsubsection{Visualization of Hidden Representations}
\label{subsubsec:visualization}

To provide an intuitive understanding of how semantic alignment refinement transforms hidden representations, we visualize the hidden states in a reduced 2D space using PCA on the Instruments dataset.

\begin{figure}[!t]
\centering
\includegraphics[width=\columnwidth]{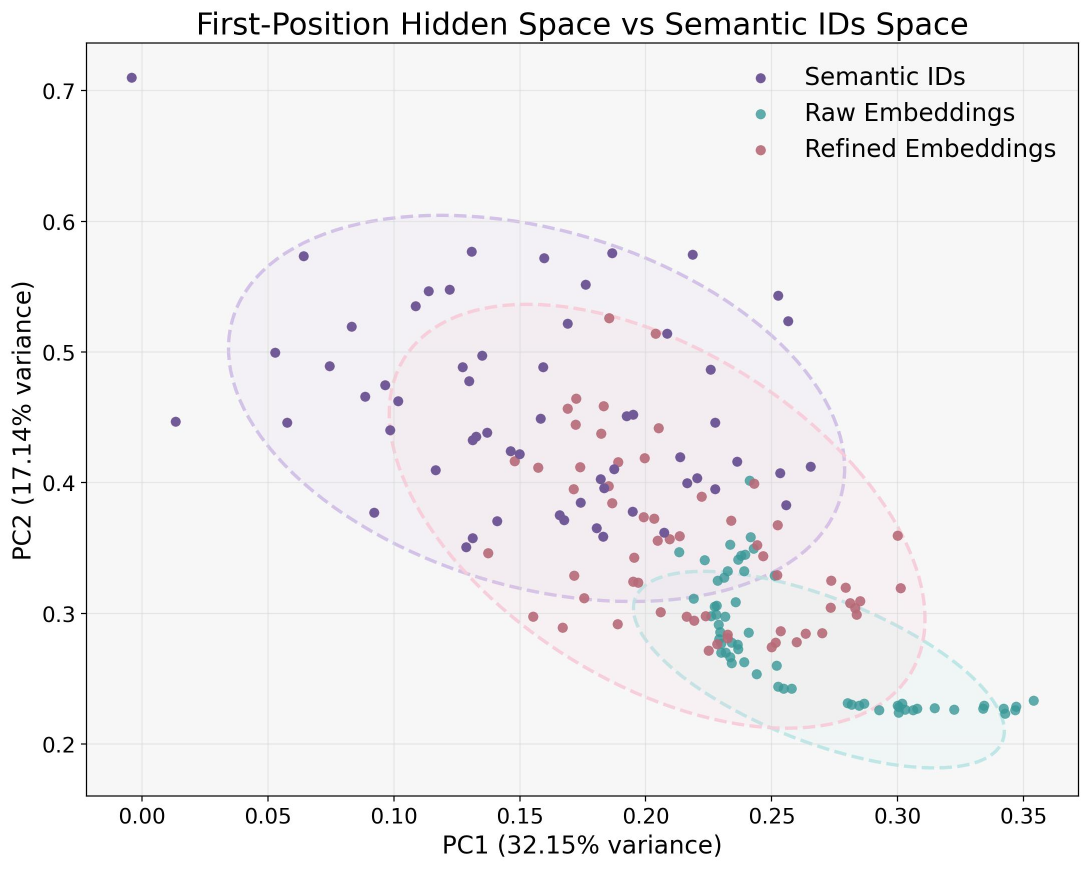}
\caption{PCA visualization of raw decoder outputs, semantically aligned outputs, and semantic ID embeddings on Instruments at the first semantic ID position.}
\label{fig:hidden_bridging}
\end{figure}

Fig.~\ref{fig:hidden_bridging} shows that semantic alignment refinement brings hidden states closer to the semantic ID space. It plots three types of points on the test set: (1) the raw decoder output $H_{\text{raw}}$ before semantic alignment refinement, (2) the refined output $H_{\text{out}}$ after semantic alignment refinement, and (3) the ground-truth semantic ID embeddings obtained from the RQ-VAE codebook. The key observation is that $H_{\text{out}}$ lies closer to the corresponding semantic ID embeddings than $H_{\text{raw}}$. This supports the role of vertical semantic alignment described in Section~\ref{subsubsec:vertical_semantic_alignment_refinement}: it maps the decoder hidden representation toward the RQ-VAE codeword space before semantic ID prediction, making the refined representation more compatible with the target codebook.

\subsubsection{Step-wise Refinement Analysis}
\label{subsubsec:refinement_dynamics}
Beyond the geometric visualization, we further examine how the effect of latent refinement varies across semantic ID positions by measuring token-level accuracy at each refinement step. This analysis is conducted using the checkpoint trained with uniform $k=2$ refinement, so that every semantic ID position receives the same number of refinement steps. This setting allows us to test whether extra refinement improves token-level accuracy at later positions. Specifically, we measure the Hit@5 of each $H_{\text{out}}$ vector, i.e., whether the top-5 nearest semantic ID tokens in the codebook include the ground-truth token at that position, under teacher forcing on the Instruments and MicroLens datasets.

\begin{figure}[!t]
\centering
\includegraphics[width=\columnwidth]{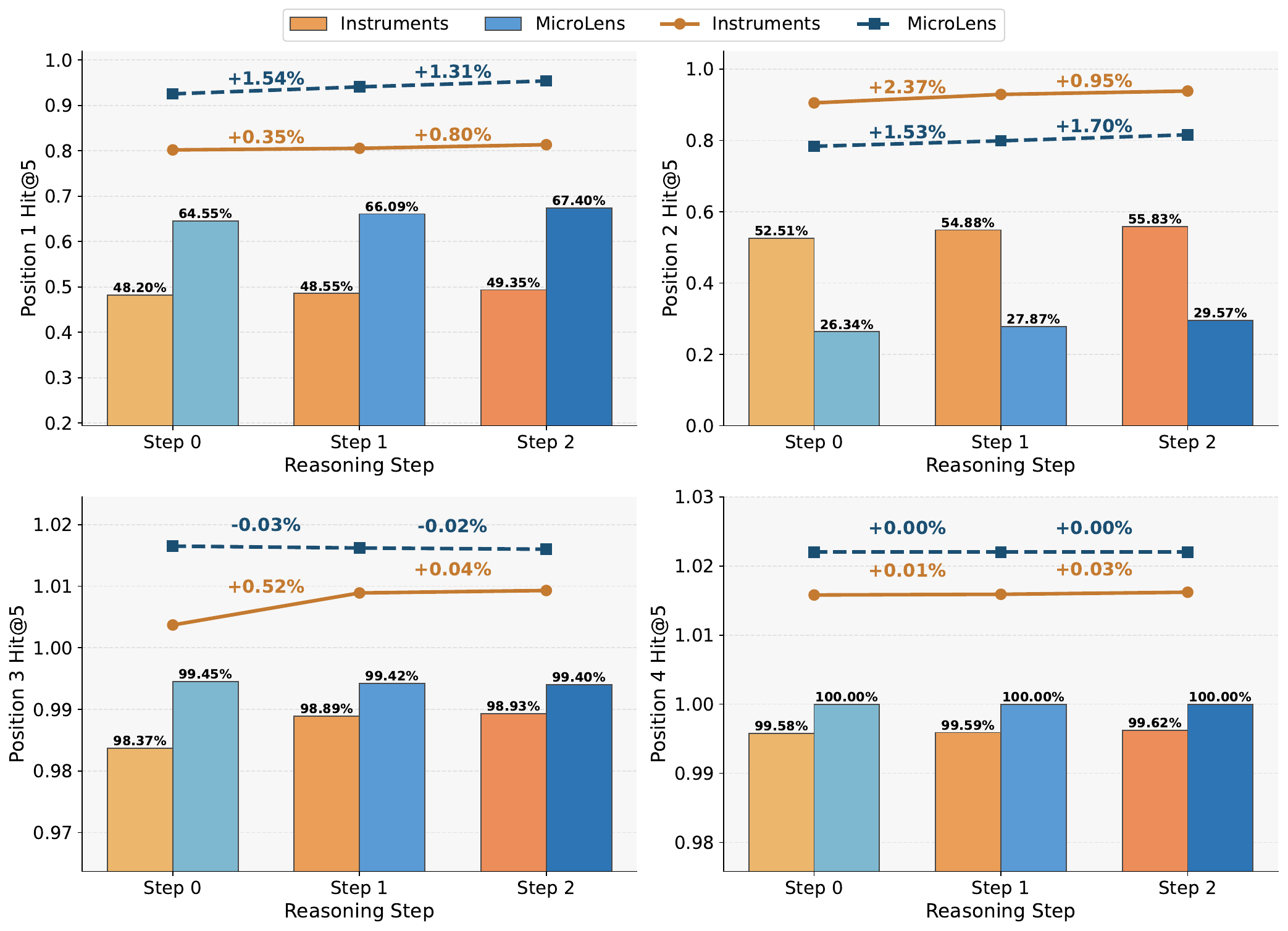}
\caption{Token-level Hit@5 of $H_{\text{out}}^{(j)}$ from Step 0 to Step 2 under teacher forcing, evaluated using the uniform $k=2$ checkpoint.}
\label{fig:stepwise_hout_hit}
\end{figure}

Fig.~\ref{fig:stepwise_hout_hit} reports the token-level Hit@5 from Step 0 to Step 2 at every position, evaluated under teacher forcing with the generation trie constraint on both the Instruments and MicroLens datasets. Several consistent patterns emerge across both datasets. At positions 1 and 2, Hit@5 improves steadily as refinement proceeds, indicating that additional refinement steps move the hidden representation closer to the ground-truth token. In contrast, positions 3 and 4 are already near-saturated at Step 0, and subsequent refinement steps yield negligible further improvement.

These results provide empirical support for the refinement gain analysis in Section~\ref{subsec:ig_analysis}. Early positions not only have higher information-gain, but also obtain higher token-level improvements from additional refinement steps, corresponding to a larger $\rho_t(k)$ in Eq.~\eqref{eq:marginal_refinement_utility}. In contrast, later positions are already near saturation, so their additional refinement utility is limited and may even decrease when unnecessary refinement disturbs reliable representations.

\section{Related Work}
\label{sec:related_work}

In this section, we review related work on generative recommendation with semantic IDs and reasoning-enhanced recommendation.

\subsection{Semantic ID-based Generative Recommendation}

\subsubsection{Generative Recommendation}
Generative recommendation reformulates recommendation as a sequence generation task, directly generating item identifiers conditioned on user behavior sequences. TIGER~\cite{rajput2023recommender} pioneers semantic ID-based generation by using RQ-VAE item codes and an encoder-decoder model to autoregressively generate the next item's semantic ID sequence. Subsequent studies further scale or extend this paradigm. HSTU~\cite{zhai2024actions} studies generative ranking from raw user behavior sequences, and OneRec~\cite{zhou2025onerec} reformulates the recommendation pipeline as autoregressive generation for industrial deployment.

\subsubsection{Semantic IDs}
A key component of generative recommendation is the design of semantic IDs, which serve as discrete tokenized item representations. Existing semantic ID construction methods can be broadly grouped into parallel and hierarchical structures. Parallel semantic IDs are often produced by product quantization~\cite{ge2013optimized}, with representative methods including VQ-Rec~\cite{hou2023learning}, RPG~\cite{hou2025generating}, and ActionPiece~\cite{hou2025actionpiece}.

Another widely used family constructs hierarchical semantic IDs through residual quantization, such as RQ-VAE~\cite{lee2022autoregressive}. Recent work improves hierarchical semantic IDs from different perspectives. PSRQ~\cite{wang2025progressive} preserves prefix semantic features to mitigate semantic drift, and HiD-VAE~\cite{fang2025hid} introduces hierarchical supervision and uniqueness regularization. LETTER~\cite{wang2024learnable} incorporates collaborative signals and mitigates code assignment bias, while TokenRec~\cite{qu2025tokenrecb}, TRM~\cite{zhao2026farewell}, and ETEGRec~\cite{liu2025generative} study tokenization designs for large-scale or end-to-end recommendation. These studies improve the quality of semantic IDs, whereas our work focuses on how latent refinement steps should be allocated across semantic ID positions during generation.

\subsection{Reasoning-Enhanced Recommendation}

\subsubsection{LLM Reasoning}
Recent advances in large language models have produced a broad family of reasoning methods. Chain-of-Thought prompting~\cite{wei2022chain} elicits intermediate natural-language reasoning steps, Tree of Thoughts~\cite{yao2023tree} explores multiple reasoning branches, and DeepSeek-R1~\cite{guo2025deepseek} shows that reinforcement learning can incentivize reasoning behaviors. Latent reasoning instead uses hidden states or repeated computation as intermediate reasoning carriers, as explored by Leap-of-Thought~\cite{talmor2020leap}, Coconut~\cite{hao2024training}, Reasoning to Learn from Latent Thoughts~\cite{ruan2025reasoning}, and Reasoning with Latent Thoughts~\cite{saunshi2025reasoning}.

\subsubsection{Reasoning for Generative Recommendation}
Recent work has introduced reasoning mechanisms into generative recommendation. R$^2$ec~\cite{you2026r$^2$ec} unifies reasoning and recommendation in a large model, while OnePiece~\cite{dai2025onepiece} and OneRec-Think~\cite{liu2025onerecthink} use context engineering or in-text reasoning scaffolds for industrial recommendation scenarios.

Latent reasoning methods enhance internal computation without generating explicit reasoning text. CARE~\cite{lin2026bringing} introduces reasoning into semantic ID-based generative recommendation from a cascaded-ranking perspective. STREAM-Rec~\cite{zhang2025slow} uses a multi-step deliberative process to improve sequential recommendation. LatentR\textsuperscript{3}~\cite{zhang2025reinforced} shifts recommendation reasoning to compact latent states and optimizes the process with reinforcement learning. SIDReasoner~\cite{he2026reasoning} studies reasoning over semantic IDs by strengthening semantic ID-language alignment and applying outcome-driven optimization.

These methods show that additional reasoning computation can improve recommendation. However, most existing approaches allocate reasoning computation uniformly or according to heuristic rules, without explicitly considering that semantic ID positions contribute unequally to item identification. In contrast, IBA studies the position-wise information structure of semantic ID generation and allocates latent refinement steps accordingly.

\section{Conclusion and Future Directions}
\label{sec:conclusion}

This paper introduces IBA, an Information-Gain Budget Allocation framework for semantic ID-based generative recommendation. IBA starts from the observation that semantic ID positions contribute unequally to item identification, and treats latent refinement steps as a limited computational resource. By combining an information-gain prior, a learned refinement gain predictor, and stable hidden-space refinement, IBA improves strong semantic ID-based generative recommendation baselines across multiple datasets.

Looking ahead, this work suggests that the position-wise information structure of semantic ID generation should be explicitly considered when allocating latent refinement steps. The current step-allocation strategy follows the observed information-gain pattern while adapting the allocation to the current user context. Future work could explore more flexible mechanisms to adjust refinement depth according to item semantics or prefix uncertainty.

\bibliographystyle{IEEEtran}
\bibliography{IEEEabrv,ref}
\end{document}